\documentclass[graybox]{svmult}
\usepackage[hyperfootnotes=true,colorlinks]{hyperref}
    	\hypersetup{colorlinks,%
    	citecolor=blue,%
    	filecolor=green,%
    	linkcolor=blue,%
    	urlcolor=blue,%
   	}

\usepackage{mathptmx}       
\usepackage{helvet}         
\usepackage{courier}        
\usepackage{type1cm}        
\usepackage{makeidx}         
\usepackage{graphicx}        
\usepackage{multicol}        
\usepackage[bottom]{footmisc}

\usepackage{amsmath,amssymb}

\makeindex             

\newcommand{\av}[1]{\left \langle #1 \right \rangle}

\newcommand{\ignore}[1]{}

\begin{document}

\title*{Cyclic structure induced by load fluctuations in adaptive transportation networks}
\author{Erik Andreas Martens and Konstantin Klemm}
\institute{Erik Andreas Martens\at Dept. of Applied Mathematics and Computer Science, Technical University of Denmark, Richard Petersens Plads, 2800 Kgs. Lyngby, \email{eama@dtu.dk}
\and Konstantin Klemm\at IFISC (CSIC-UIB), Campus Universitat de les Illes Balears, E-07122 Palma de Mallorca, Spain \email{klemm@ifisc.uib-csic.es}}
%
%
\maketitle

\abstract{\ \ 
Transport networks are crucial to the functioning of natural systems and technological infrastructures. 
For flow networks in many scenarios, such as rivers or blood vessels, acyclic networks (i.e., trees) are optimal
structures when assuming time-independent in- and outflow. Dropping this assumption, fluctuations
of net flow at source and/or sink nodes may render the pure tree solutions unstable even under a simple
local adaptation rule for conductances. Here, we consider tree-like networks under the influence of spatially heterogeneous distribution of fluctuations, where the root of the tree is supplied by a constant source and the leaves at the bottom are equipped with sinks with fluctuating loads.  
We find that the network divides into two regions characterized by tree-like motifs and stable cycles. The cycles emerge through transcritical bifurcations at a critical amplitude of fluctuation.  For a simple network structure, depending on parameters defining the local adaptation, cycles first appear close to the leaves (or root) and then appear closer towards the root (or the leaves). The interaction between topology and dynamics gives rise to complex feedback mechanisms with many open questions in the theory of network dynamics. 
A general understanding of the dynamics in adaptive transport networks is essential in the study of mammalian vasculature, and adaptive transport networks may find technological applications in self-organizing piping systems.
}
%

\section{Introduction}
\label{sec:intro}

Network modeling deals with the rules for establishing and removing connections between the entities that make up a system. For instance, social networks display a much larger amount of triangles than expected under entirely random wiring. A possible explanation is that nodes (i.e., persons) are more likely to introduce their already existing friends to each other~\cite{Davidsen2002}. Similarly, for biological networks, a simple network growth rule of node copying and random perturbation of edges mimics genome duplication and thereby reproduces statistical features of protein interaction networks~\cite{Sole2002,Ispolatov2005}.

More recent models of adaptive networks involve bidirectional dependence between a dynamics in the networked system and the dynamic modification of its link structure~\cite{Gross2008,Herrera2011,porter2016dynamical}. Here we study this dependence specifically for the case of a network for transport and distribution, motivated by the vascular (blood circulatory) system in higher animals. 
This system fulfills the task of transport from one central source (heart / lung) to spatially distributed sinks. Assuming a constant in-flow from the source and a constant outflow into sinks, the optimal distribution system in terms of energy consumption is a tree~\cite{Kantorovich1942translocation,Villani2003}, i.e.,\ a cycle-free connected network. When load at the sinks fluctuates, however, networks involving cycles become optimal as shown by Corson \cite{Corson2010} and Katifori with colleagues \cite{Katifori2010}.

Here we combine this insight with local adaptation~\cite{HuCai2013,MartensKlemm2017} rather than global optimization. Indeed, such models are relevant in vascular physiology, where arterioles adapt their diameter and wall thickness on time scales from seconds over days to months in dependence on local flow variables including pressure and flow shear~\cite{Jacobsen2008,Jacobsen2009}.
The conductances of the flow network self-organize towards balanced pressure fluctuations. We observe that cycles form only when the amplitude of load fluctuations exceeds a threshold. With the source placed at the top and all sinks in the bottom layer of a hierarchical network, as illustrated in Fig.~\ref{fig:network_topologies}, we show that cycle formation is localized: depending on a parameter of the adaptation rule, there is a transition between cycle formation close to the source or cycle formation close to the sinks.

\section{Model}
\label{sec:model}

$ V $ denotes the set of nodes of the network with $ N = |V|<\infty $ and $ A \subseteq N \times N $ the set of edges. The edges are bidirectional, so $(i,j) \in A$ implies $(j,i) \in A$. Each node is assigned a pressure $ p_i $. 
The edge flow is $ f_{ij}>0 $ from node $ i $ to $ j $.
Furthermore we assume that the network is resistive and linear, i.e., it is Ohmian with $ f_{ij} = C_{ij} (p_i-p_j) $, where an edge carries the property of a conductance between nodes $ i $ and $ j$ with $ C_{ij}=C_{ji} > 0 $ only if $ (i,j) \in A$; otherwise $ C_{ij}=C_{ji} = 0$.
\begin{figure}[htp!]
 \centering
 \includegraphics[width=0.7\textwidth]{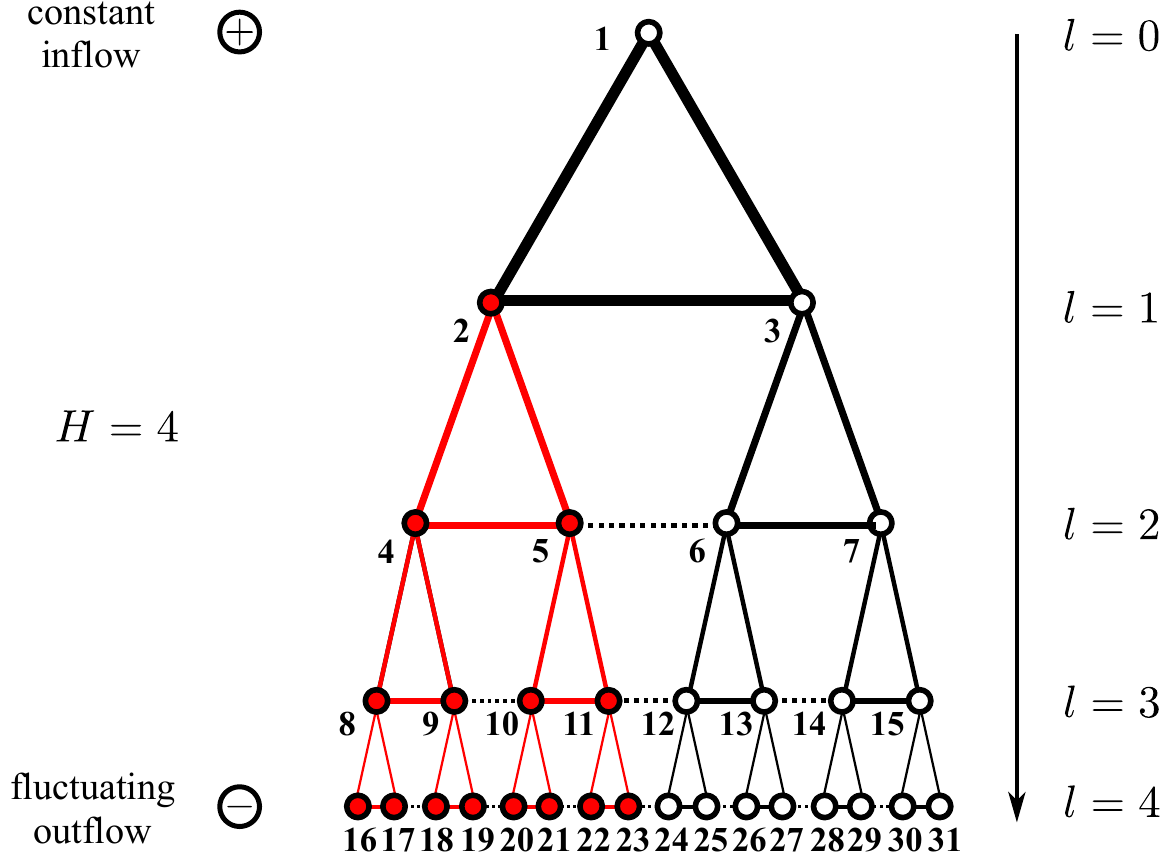}
 \caption{\label{fig:network_topologies} 
 'Simple Augmented Trees' (SAT) are tree-like structures with \emph{cross-edges} (full horizontal edges) that connect only nodes in each minimal subtree (e.g., edges highlighted in red), thus forming a \emph{'tree of triangles'}; 
 the 'Fully Augmented Tree' (FAT) connects all nodes within one tree level $l$ by a path.  The branching level in the cut tree is denoted by $l$.
 Cross-edges introduce cycles to the network. The root of the tree (top) and the fluctuating sinks in the leaves of the tree (bottom) drive the flow. Cross-edges emerge depending on the strength of the fluctuation in the leaves. For SATs, the dynamics in a triangular submotif at level $l$ (red edges) will only depend on downstream fluctuations (red nodes). 
 }
\end{figure}

Here, we study tree-like networks of height $H$ as illustrated in Fig.~\ref{fig:network_topologies}, with cross-edges on every branching level, $l=0,\ldots, H$. Cross-edges lead to cyclic structure. A cycle (red triangle) is a connected subnetwork of $m$ nodes such that each node has exactly two neighbours. We focus on two types of networks: the \emph{simply augmented tree} (SAT), where cross-edges only form triangular submotifs (i.e., excluding dotted cross-edges) and \emph{fully augmented tree} (FAT) where all displayed cross-edges are allowed.	

To model \emph{sources and sinks} in the network, we include non-zero nodal flows $h_i$. 
A proper subset $S$ of the node set $V$ is chosen as the set of sink nodes. The set $S$ is time-independent and typically comprises the most peripheral nodes, where capillaries connect to the vein network. With a tree structure underlying the network, $S$ is chosen as the set of leaves of the tree. Focusing on the networks based on symmetric trees of height $H$ (cf.~Figure~\ref{fig:network_topologies}), we have the $|S| = n = 2^H$ leaves as sink nodes. For each sink node $i \in S$, the nodal flow $h_i(t)$ is non-positive at all times $t\in \mathbb{R}$. A single node in $V \setminus S$ is chosen as the source node and 
indexed as node $1$ for simplicity. For the networks based on symmetric trees, the source node is the root of the tree. The source node has a positive nodal flow $h_1(t)=1$ for all $t\in \mathbb{R}$. For all other nodes $j \in V \setminus (S \cup \{1\})$, we set $h_j(t)=0$ for all $t\in \mathbb{R}$. Mass balance requires that $\sum_{k \in V} h_k(t) = 0 $ for all $t\in \mathbb{R}$.

Assuming that the accumulation rate of fluid at any node is nearly instantaneous, or that vessels are inelastic, the nodal accumulation rate becomes negligible~\cite{Reichholdt2009,MartensKlemm2017}, and we may express
mass  balance by invoking Kirchhoff's first law,
\begin{equation}
 \sum_{j} C_{ij}({p}_i-{p}_j) =  {h}_i,\
\end{equation}
which is re-written in vector/matrix notation by defining the nodal flow  ${\mathbf{ h }} := ({h}_i)_{i \in V}$ and the Kirchhoff matrix ${\mathbf{K}}=({K}_{ij})_{i,j\in V}$ with $K_{ij}:=(\delta_{ij}\sum_{j^\prime} C_{i{j^\prime}})-C_{ij}$,   
\begin{equation}\label{eq:Kirchhoff}
 {\mathbf{K}} \cdot {\mathbf{p}} = {\mathbf{h}}\
\end{equation}f
which is solved for ${\mathbf{ p }} := ({p}_i)_{i \in V}$.

To impose \emph{adaptive dynamics to the network}, we postulate the generic ad-hoc law for the conductances~\cite{MartensKlemm2017}:
\begin{equation}\label{eq:adaptation}
  \frac{d}{dt} C_{ij} = \alpha_1 C_{ij} |p_j-p_i|^\gamma - \alpha_2 C_{ij}.\
\end{equation}
Thus, the first term on the right hand side induces growth proportional to the power dissipated along the edge, thus mitigating rising pressure differences by increasing the conductance along the edge. The network adapts towards minimizing power consumption. The last term prevents unlimited growth of the conductances. 

Rescaling variables with $ C_{ij}':=h_1^{-1}(\alpha_2/\alpha_1)^{1/\gamma}\, C_{ij} $ and $ p_i':=(\alpha_1/\alpha_2)^{1/\gamma} \, p_i $, $h':=h/h_1$ (so that $h_1'=1$), $t':=\alpha_2 t$,
the resulting dimensionless model reads 
\begin{eqnarray}\label{eq:goveqns}
 \frac{d}{dt'} C_{ij}'(t) &=& C_{ij}'(t)[ |p_j'(t)-p_i'(t)|^\gamma - 1],\\\label{eq:Kirchhoff2}
 \mathbf{K}'(t) \cdot \mathbf{p}'(t) &=& 	 \mathbf{h'}(t).\
\end{eqnarray}
where we drop the primes and omit the argument $t$ from now on. Note that the solvability condition of the Kirchhoff equation~\cite{MartensKlemm2017}, $ \sum_{j} h_j'(t)=0 $, follows from Fredholm's 
alternative and has the physical interpretation of mass conservation.


We consider sinks with varying load, compliant with $ \sum_{k \in V} h_k(t) = 0 $. For each time $t$, there is a sink node $s \in S$ so that the nodal flow $\mathbf{h}(t)$ is the
vector $\mathbf{g}^{(s)}$ with components 
\begin{equation} \label{eq:stochfluc}
  g_i^{(s)} = \begin{cases}
  +1                                        & \text{if } i \text{ is the source (root) node.}\\
  - \dfrac{1}{n} - \dfrac{a}{\sqrt{2}}=:h_- & \text{if } i=s \\
  - \dfrac{1}{n} + \dfrac{1}{n-1}\dfrac{a}{\sqrt{2}}=:h_+  & \text{if } i \in S\setminus\{s\} \\
  0                                                        & \text{otherwise.}
\end{cases}
\end{equation}
This reflects the situation where at each point $t$ in time, one of the sinks has higher load ($h_-$) than the others (having load $h_+$). 
Independent of time, the source (root node) has an inflow of $+1$. All other nodes have neither in- nor outflow. The driving amplitude $a$ is a parameter of the model,
obeying $a \in [0,a_\text{max}]$ with $a_\text{max} = \sqrt{2}(|S|-1)/|S|$. This ensures $h_- \le h_+ \le 0$ meaning sink nodes actually behave as sinks at all times.
In the extreme case $a=a_\text{max}$, only the sink with the higher load is on ($h_-=-1$) while all others are off ($h_+=0$). This case reproduces the single moving sink as employed 
by Katifori {\it et al.}~\cite{Katifori2010}.

Following~\cite{MartensKlemm2017}, we assume that (i) all sinks are in the high load state for the same fraction of time and (ii) the variation of sink load occurs on a time scale faster than the adaptation of the conductances.
Thus, equations (\ref {eq:goveqns}) are effectively equivalent to an averaged form as follows,
\begin{equation}\label{eq:goveqns_av}
 \frac{d}{dt} C_{ij}(t) = C_{ij}(t)[ \av{ |p_j(t)-p_i(t)|^\gamma } - 1],\\
\end{equation}
where time $t$ corresponds to the slow time and the average $\av{.}$ is taken uniformly over the $n=|S|$ assignments of high load sinks,
\begin{equation}
\av{ |p_j(t)-p_i(t)|^\gamma } := n^{-1} \sum_{s \in S} |p^{(s)}_j(t)-p^{(s)}_i(t)|^\gamma
\end{equation}
and
\begin{equation}
\mathbf{K}(t) \cdot \mathbf{p}^{(s)}(t) = \mathbf{g}^{(s)}~.
\end{equation}
For simplicity, we drop the averaging brackets $\av{\cdot}$ from now on.

\section{Analysis}\label{sec:analysis}

\subsection{The case of exponent $\gamma =2$}

\begin{figure}[htp!]
\sidecaption
\includegraphics[width=\textwidth]{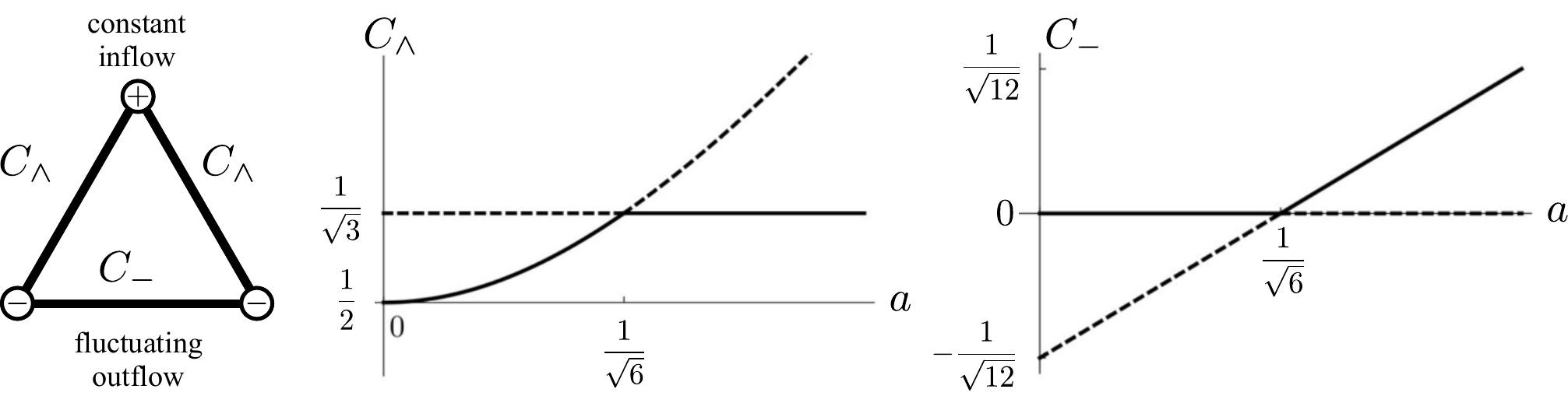}
\caption{
Solution branches for the triangular graph with $H=1$ undergo a sub-critical bifurcation where the conductance of the cross-edge, $C_-$, becomes non-zero.
}
\label{fig:solution_triangular_nw}       
\end{figure}

Our previous analysis~\cite{MartensKlemm2017} of the model concentrated on the case $\gamma=2$.  With this choice, the growth of an edge with conductance $C_{ij}$ in~\eqref{eq:adaptation} is proportional to the power (dissipated energy per time) over this same edge. For the height $H=1$, the simply/fully augmented tree becomes a triangle of nodes $V=\{1,2,3\}$ as shown in Fig.~\ref{fig:solution_triangular_nw}. The two sink nodes, indexed 2 and 3 are connected to each other with a conductance $C_-:=C_{23}$, and each of them also to the source node, indexed 1, with the symmetric conductance $C_\wedge:=C_{12}=C_{13}=$ of the cut-edges. We are interested in (slow time) stationary solutions  of the model driven with load fluctuations of amplitude $a$ as a single parameter. 
The stationary conductance $C_{-}$ of the cross-edge connecting the two sinks is of particular interest. 
The solutions undergo a transcritical bifurcation at parameter value $a_c=1/ \sqrt{6} \approx 0.408$. 
For sub-critical fluctuation amplitude, $a < a_c$, the uniquely stable solution branch has $C_{-} = 0$ for the cross edge and $C_\wedge>0$ for the cut edge which grows monotonically with $a$. 
For super-critical amplitude, $a > a_c$, the stable branch has a positive conductance for the cross-edge, $C_{-}>0$, which grows almost linearly with increasing amplitude $a$, while the cut-edge stays exactly constant. Thus, the cross-edge short circuits fluctuations so that the two cut-edges may stay constant.

In simply augmented trees with more layers ($H>1$, see Fig.~\ref{fig:network_topologies}), all cross-edge conductances undergo transcritical bifurcations from zero to non-zero as well~\cite{MartensKlemm2017,MartensKlemm2018}. The parameter value $a_c^{(l,H)}$ at the transition depends both on the level $l$ of the cross-edge and the system height $H$. In a given system with height $H$, as the fluctuation amplitude $a$ increases, cross-edges at the sink nodes first undergo a transcritical transition from zero to positive conductance. As the amplitude $a$ is increased further, cross-edges at the next level become non-zero, thus following a strict ordering $a_c^{(l+1,H)} < a_c^{(l,H)}$ for all $l \in \{1,\dots,H-1\}$. The resulting ordering is qualitatively similar as illustrated in Fig.~\ref{fig:nipcsg_0} (c) where $\gamma=1.5$. The strict ordering in terms of level $l$ may be directly linked to the topology of the SAT, which (including its cross edges) forms a tree of triangular submotifs~\cite{MartensKlemm2018}.

\subsection{Effect of varying exponent $\gamma$}

\begin{figure}[htp!]
 \centering
 \includegraphics[width=0.9\textwidth]{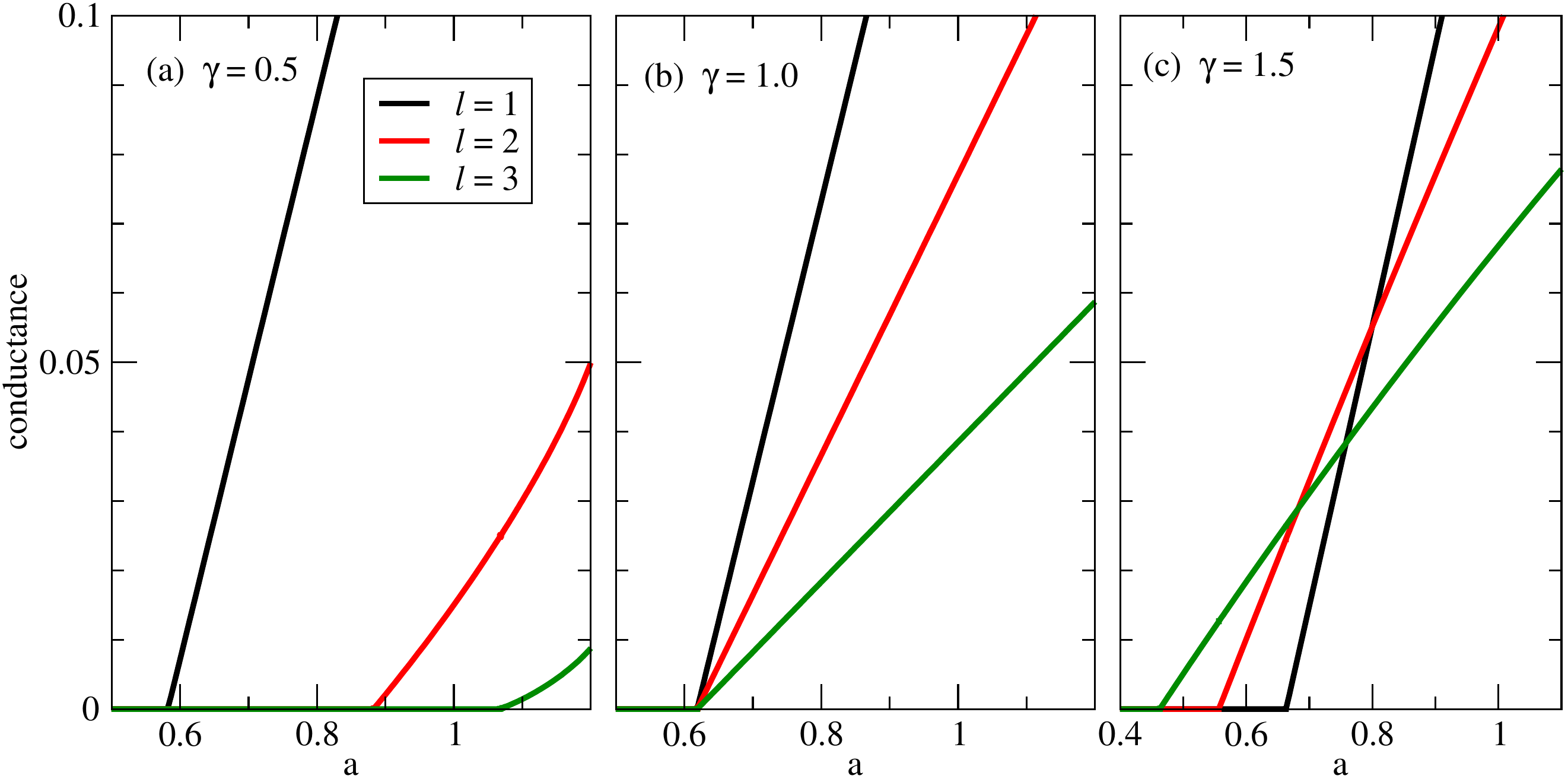}
 \caption{\label{fig:nipcsg_0} 
 Conductances of cross-edges in simply augmented trees of height $H=3$ as a function of sink fluctuation parameter. The three panels distinguish exponent values ({\bf{a}}) $\gamma=0.5$, ({\bf{b}}) $\gamma=1.0$ and ({\bf{c}}) $\gamma=1.5$ which affects the transition order for different levels $l$.
 }
\end{figure}

Fig.~\ref{fig:nipcsg_0} shows the influence of $\gamma$ on the transcritical bifurcations in the simply augmented tree of height $H=3$. For $\gamma=1.5$, the transitions occur in the same ordering as known for $\gamma=2$, i.e. from sink node level (here $l=3$) towards source node level. For $\gamma=0.5$, the order of transitions is reversed, while $\gamma=1$ has all transitions at the same parameter value.

\begin{figure}[htp!]
 \centering
 \includegraphics[width=0.9\textwidth]{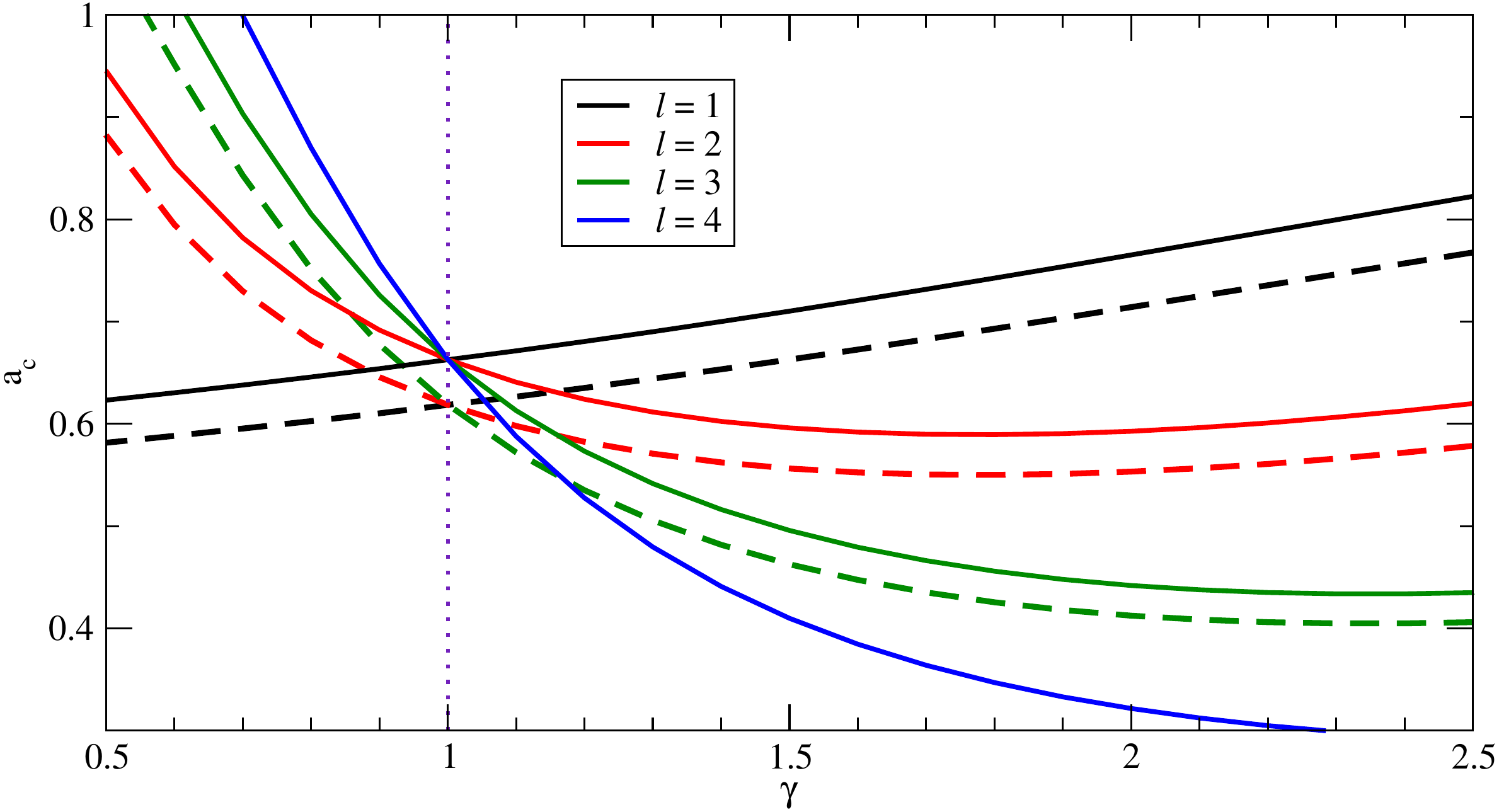}
 \caption{\label{fig:ac_gamma_7}
 Critical fluctuation amplitude values $a_c^{(l,H)}$ for cross-edges in simply augmented trees of height $H=3$ (dashed lines) and $H=4$ (solid lines), as a function of exponent $\gamma$.
 }
\end{figure}

Fig.~\ref{fig:ac_gamma_7} shows the $\gamma$-dependence of critical values of parameter $a$ for cross-edges at different levels $l$ in systems of heights $H=3$ and $H=4$. This confirms, here numerically for the provided cases, the change of behaviour at $\gamma=1$. For $\gamma <1$, the onset of cross-edge conductances happens first close to the source and moves downward in the system. For $\gamma>1$ this order is reversed.

In fully augmented trees, the ordering of the critical amplitudes, $a_c^{(l,H)}$, follow a more complex pattern and is subject to further study~\cite{MartensKlemm2018}.

\section{Conclusion and Discussion}
\label{sec:discussion}

We have studied conditions for the presence or absence of cycle forming edges in a model of vascular networks under load fluctuations. Variation of the exponent $\gamma$ in the local pressure dependence of conductance adaptation changes the order by which cycles arise in the simply augmented trees (Fig.~\ref{fig:network_topologies}), either first close to the sink or close to the root. For network structures with less symmetry, preliminary analysis finds a more complex ordering sequence. This reflects 
that, generally, the interaction between topology and dynamics gives rise to complex feedback mechanisms posing open questions in the theory of network dynamics.
Future work on this model ought to include a comparison to empirical vascular networks. Both the network structures themselves and measurements on the flow through branches of the network are becoming available~\cite{Blinder2013,Poelma2017,Alim2018}. 

In view of data and for better alignment with real structures, the tree-like networks of this model may be augmented further. In a first step, one may include further edges inside a level as indicated by the dashed lines in Fig.~\ref{fig:network_topologies}. For this fully augmented tree, preliminary analysis \cite{MartensKlemm2017} has shown that cross-edges become conductive in a pattern similar to that of the  simply augmented tree at exponent $\gamma=2$. Further results will be reported elsewhere~\cite{MartensKlemm2018}.  A complete quantitative understanding of the transitions in this system would bring us closer to a general theory for the emergence of cycles in transport networks in the presence of fluctuations. 

A further interesting step would be to abandon imposed network structures altogether and cast the adaptation dynamics into real two- or three-dimensional space. Having both conductance $c$ and pressure $p$ as scalar fields, adaptation of conductances can be described \cite{MartensKlemm2018} by an equation as
\begin{equation}
\partial_t c(x,t) = c(x,t) [ (\nabla p(x,t))^2 -1 ]~
\end{equation}
as a proposal for the real-space analog of the adaptation rule in Eq.~\eqref{eq:goveqns}.

Beyond modeling the self-organization of transport networks in nature, this branch of research has bearings also in technical applications. {\em Programmable materials} is a branch of technology to generate complex objects by self-assembly of their suitably programmed constituents \cite{Campbell2014}. Recent ideas and advances point in the direction of evolutionary materials that are capable of self-repair and adaptation to changing environmental conditions \cite{Papadopoulou2017}. A pertinent example are urban water-supply systems where pipes self-adapt the flow capacity in response to local demand fluctuations
in a city with evolving population density \cite{Campbell2014}.


%
\begin{acknowledgement}
EAM and KK acknowledge travel funding from Action CA15109, European Cooperation for Statistics of Network Data Science (COSTNET).
KK acknowledges funding from MINECO through the Ram\'{o}n y Cajal program and through project SPASIMM, FIS2016-80067-P (AEI/FEDER, EU). We thank J. C. Brings Jacobsen for helpful discussions on circulatory physiology and E. Katifori on adaptive networks.
\end{acknowledgement}

\bibliographystyle{unsrt}
\bibliography{references} 

\end{document}